# On the Sum Capacity of A Class of Cyclically Symmetric Deterministic Interference Channels

Bernd Bandemer, Gonzalo Vazquez-Vilar, and Abbas El Gamal
Stanford University, Information Systems Laboratory
350 Serra Mall, Stanford, CA 94305, USA
email: {bandemer, gvazquez}@stanford.edu, abbas@ee.stanford.edu

***Abstract*— Certain deterministic interference channels have been shown to accurately model Gaussian interference channels in the asymptotic low-noise regime. Motivated by this correspondence, we investigate a $K$ user-pair, cyclically symmetric, deterministic interference channel in which each receiver experiences interference only from its neighboring transmitters (Wyner model). We establish the sum capacity for a large set of channel parameters, thus generalizing previous results for the 2-pair case.**

## I. Introduction

The Gaussian interference channel (G-IC) is one of the most important and practically relevant models in multiple user information theory. Although the capacity region of this channel is not known in general, significant progress has been made recently toward finding capacity under weak interference [1]–[3] and bounds that are provably close to capacity [4]–[6]. In [4], the capacity region for the two user-pair G-IC is established to within one bit using new outer bounds and a simplified Han-Kobayashi achievability scheme. The same asymptotic result is derived in [5] by making a correspondence between the G-IC in low-noise regime and a class of deterministic, finite-field interference channels [7]. Some progress toward generalizing this result to more than two user-pairs has been made in [6], where the solution is found for the fully symmetric case.

Motivated by these recent results, we consider a class of $K$ user-pair ($K \geq 3$), cyclically symmetric, deterministic, finite-field interference channels in which each receiver experiences interference only from its two nearest neighbors as in the Wyner model [8]. We determine the sum capacity of this channel for a wide range of interference parameters. Because of symmetry in the channel and in the data rates, it suffices to consider the $K = 3$ case depicted in Fig. 1. We focus our discussion on this case while keeping in mind that all results generalize immediately to the $K$ user-pair Wyner case.

## II. Channel definition

Let $\mathbb{F}_2$ denote the binary finite field and let $\mathbf{I}$ be the identity matrix. The zeropadding operator $\mathbf{Z} \in \mathbb{F}_2^{2N \times N}$ is

We gratefully acknowledge the contribution of Dr. Aydin Sezgin, who originally suggested the channel that we consider in this work. Bernd Bandemer is supported by an Eric and Illeana Benhamou Stanford Graduate Fellowship and US Army grant W911NF-07-2-0027-1. Gonzalo Vazquez-Vilar is supported by Fundacion Pedro-Barrie de la Maza Graduate Scholarship.

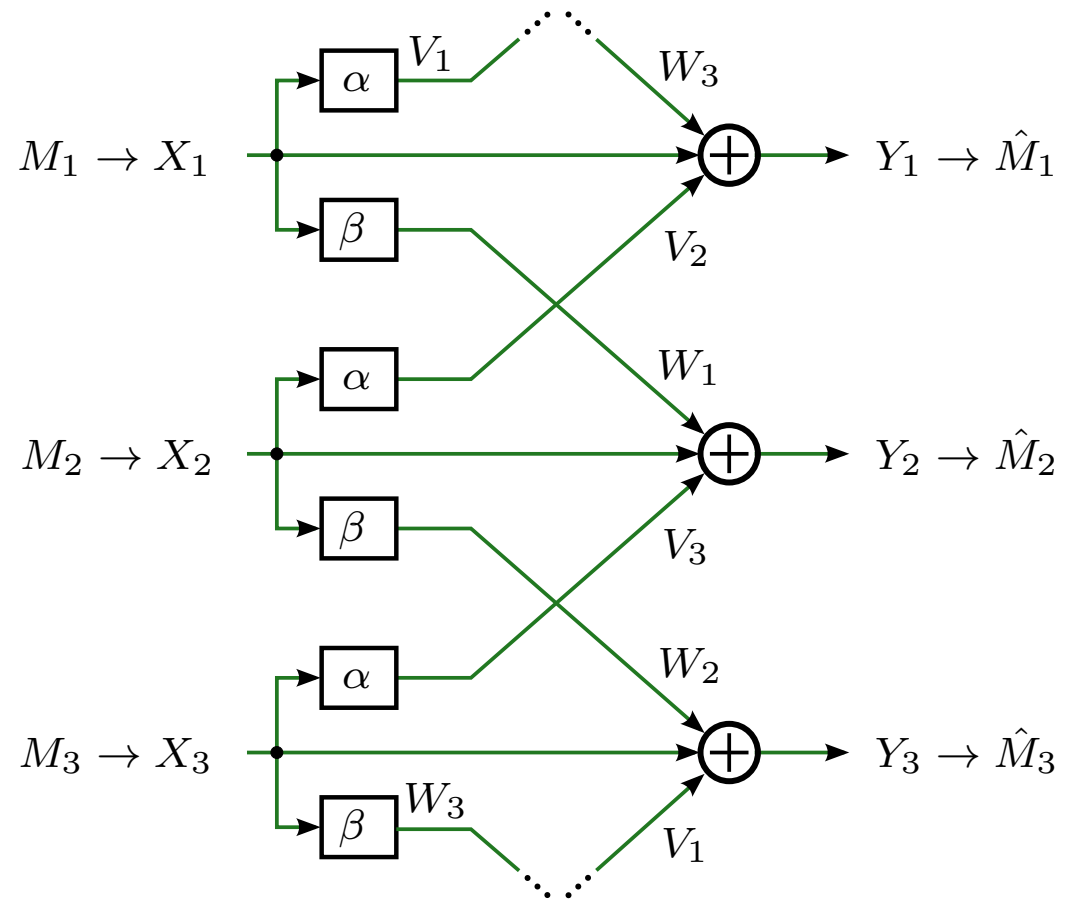

**Fig. 1.** Cyclically symmetric deterministic interference channel, $K = 3$.

defined as $\mathbf{Z} = [\mathbf{0}_{N\times N}, \mathbf{I}_N]^T$. Further, let $\mathbf{U}, \mathbf{D} \in \mathbb{F}_2^{2N\times 2N}$ be the upshift and downshift matrix, respectively, such that $\mathbf{U}[x_1, x_2, \ldots, x_{2N-1}, x_{2N}]^\mathrm{T} = [x_2, x_3, \ldots, x_{2N}, 0]^\mathrm{T}$ and $\mathbf{D}[x_1, x_2, \ldots, x_{2N-1}, x_{2N}]^\mathrm{T} = [0, x_1, \ldots, x_{2N-2}, x_{2N-1}]^\mathrm{T}$. We also use the standard notation $A^n = (A_1, A_2, \ldots, A_n)$.

We refer to a $K$ user-pair interference channel as *cyclically symmetric* if the channel is invariant to cyclic relabeling of the pairs, i.e., renaming $i$ as $i+1$ for $i < K$, and $K$ as 1.

We investigate the class of cyclically symmetric, deterministic, finite-field interference channels with $K = 3$ user-pairs depicted in Fig. 1. The channel is stationary and memoryless across multiple channel uses. The channel inputs are $X_1, X_2, X_3 \in \mathbb{F}_2^N$ and its outputs are $Y_1, Y_2, Y_3 \in \mathbb{F}_2^{2N}$, where $N$ is the number of input bit pipes at each sender. The outputs of the channel are given by

$$Y_1 = \mathbf{Z}X_1 + V_2 + W_3,$$
$$Y_2 = \mathbf{Z}X_2 + V_3 + W_1,$$
$$Y_3 = \mathbf{Z}X_3 + V_1 + W_2,$$

where $+$ is the modulo-2 addition operator, and $V_k = \mathbf{U}^{(\alpha-1)N}\mathbf{Z}X_k$, $W_k = \mathbf{D}^{(1-\beta)N}\mathbf{Z}X_k$ for every $k$.

The channel is parameterized by the triple $(N, \alpha, \beta)$, which we constrain to $\alpha \in [1, 2]$, $\beta \in [0, 1]$, and $\alpha N, \beta N \in \mathbb{Z}$. The parameters $\alpha$ and $\beta$ characterize the amount of up/downshift on the cross links and thus loosely correspond to channel gains. Since by the definition of our channel, for each user-pair there is always exactly one interferer being up-shifted and one being down-shifted, our channel is a special case of the class of cyclically symmetric, deterministic, finite field, Wyner connected channels. Note that the up-shifted $V_k$ retains the complete information of $X_k$, while the down-shifted $W_k$ incurs clipping at the low end of the vector.

Transmitter $k \in \{1, 2, 3\}$ wishes to convey an independent message $M_k$ at data rate $R_k$ to its corresponding receiver. We define a $(2^{nR_1}, 2^{nR_2}, 2^{nR_3}, n)$ code, probability of error, and achievability of a given rate triple $(R_1, R_2, R_3)$ in the standard way [9]. The capacity region $\mathcal{C}$ of the channel is the closure of the set of all achievable rate triples. Define the sum capacity as $R_\Sigma := \sup\{R_1 + R_2 + R_3 \mid (R_1, R_2, R_3) \in \mathcal{C}\}$ and the symmetric capacity as $R_{\text{sym}} := \sup\{R \mid (R, R, R) \in \mathcal{C}\}$. By symmetry of the channel and convexity of the capacity region, $R_\Sigma = 3R_{\text{sym}}$. Furthermore, define the *symmetric generalized degrees of freedom* $d_{\text{sym}} := R_{\text{sym}}/N$, i.e., the symmetric capacity normalized with respect to the interference-free case.

Before we state our main result, define the function

$$\mathcal{V}(x) := \frac{1+|x-1|}{2} = \begin{cases} \frac{x}{2} & \text{if } x \geq 1 \\ 1 - \frac{x}{2} & \text{if } x < 1. \end{cases}$$

*Remark 1:* This definition is useful in the context of deterministic finite-field interference channels. For example, the symmetric generalized degrees of freedom of the two user-pair symmetric deterministic interference channel with parameters $(N, \alpha)$, with $\alpha \in [0, \infty)$, are known [4], [5] to be

$$d_{\text{sym}} = \min\{1, \mathcal{V}(\alpha), \mathcal{V}(2\alpha)\}.$$

## III. Main Result

Our main result establishes the symmetric generalized degrees of freedom for the class of interference channels defined above for a large set of $(\alpha, \beta)$ parameters.

*Theorem 1:* The symmetric generalized degrees of freedom of the class of three user-pair, cyclically symmetric, deterministic, finite field interference channel with parameters $(\alpha, \beta) \in [1, 2] \times [0, 1]$, where $\alpha \geq 2\beta$ or $\alpha \geq \frac{\beta}{2} + 1$, is

$$d_{\text{sym}} = \min\{1, \mathcal{V}(\alpha), \mathcal{V}(\beta), \mathcal{V}(2\beta), \mathcal{V}(\alpha - \beta)\}.$$

Fig. 2 depicts our result. The claimed $d_{\text{sym}}$ is piecewise linear in $(\alpha, \beta)$, and the figure shows the linear regions in the parameter plane with their respective minimum and maximum values of $d_{\text{sym}}$. Some of the linear pieces are subdivided (for example, "Ea" and "Eb") to indicate that different achievability schemes are needed even within a single linear piece (see Section V).

*Remark 2:* The theorem implies that $d_{\text{sym}}$ is independent of $N$. For fixed $\alpha$ and $\beta$, all valid values of $N$ (satisfying $\alpha N, \beta N \in \mathbb{Z}$) yield the same $d_{\text{sym}}$.

## IV. Converse Proof

The upper bounds $1$, $\mathcal{V}(\alpha)$, $\mathcal{V}(\beta)$, and $\mathcal{V}(2\beta)$ follow in a straightforward way from the known degree of freedom result of the two user-pair case [4], [10]. This can be shown by giving the complete signal $X_k^n$ of one of the interferers as genie information to the receivers, thus effectively degenerating the three user-pair case to the two-pair case.

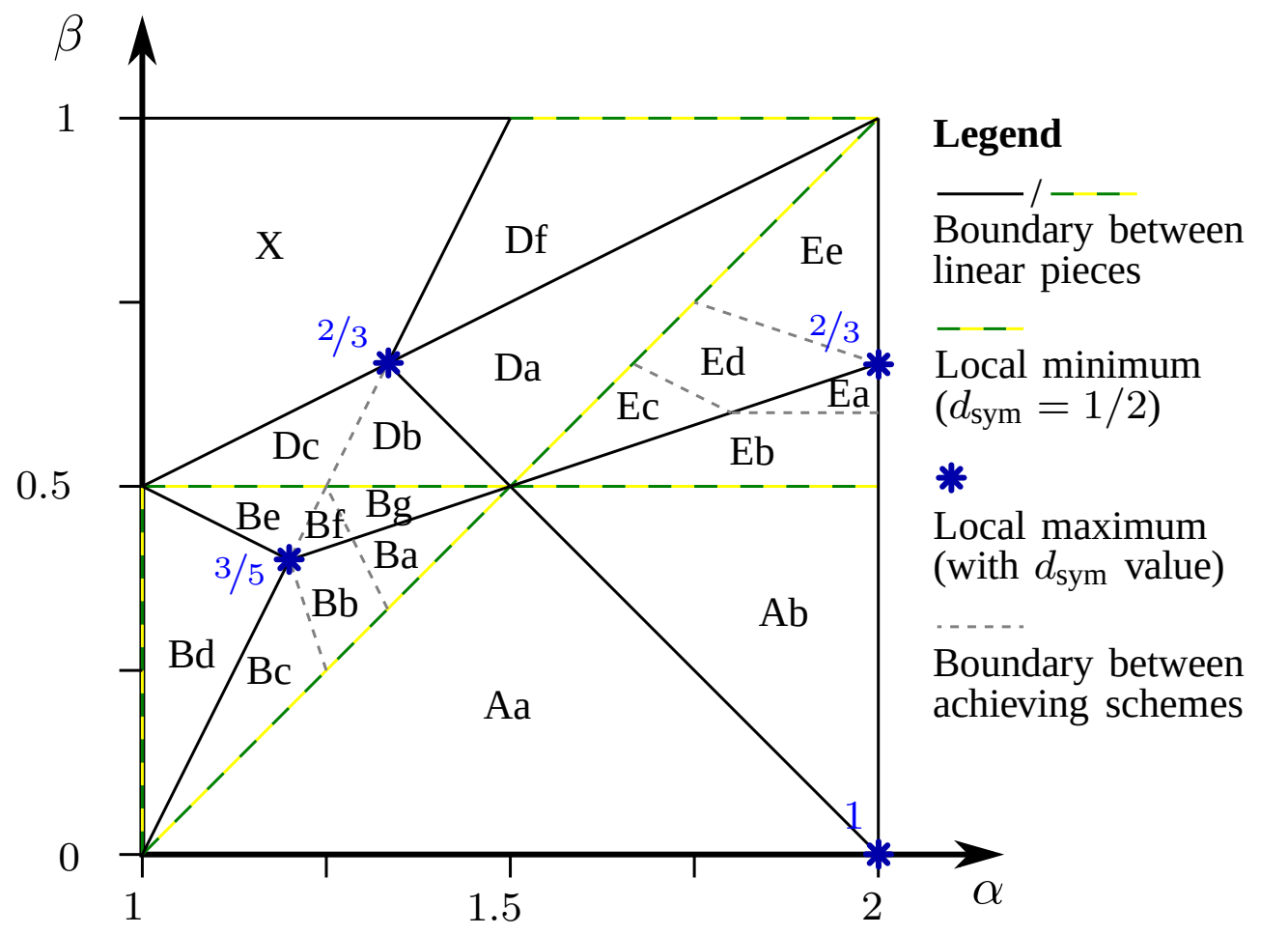

**Fig. 2.** Illustration of Theorem 1 in the $(\alpha, \beta)$ parameter plane. The result applies everywhere except in region "X".

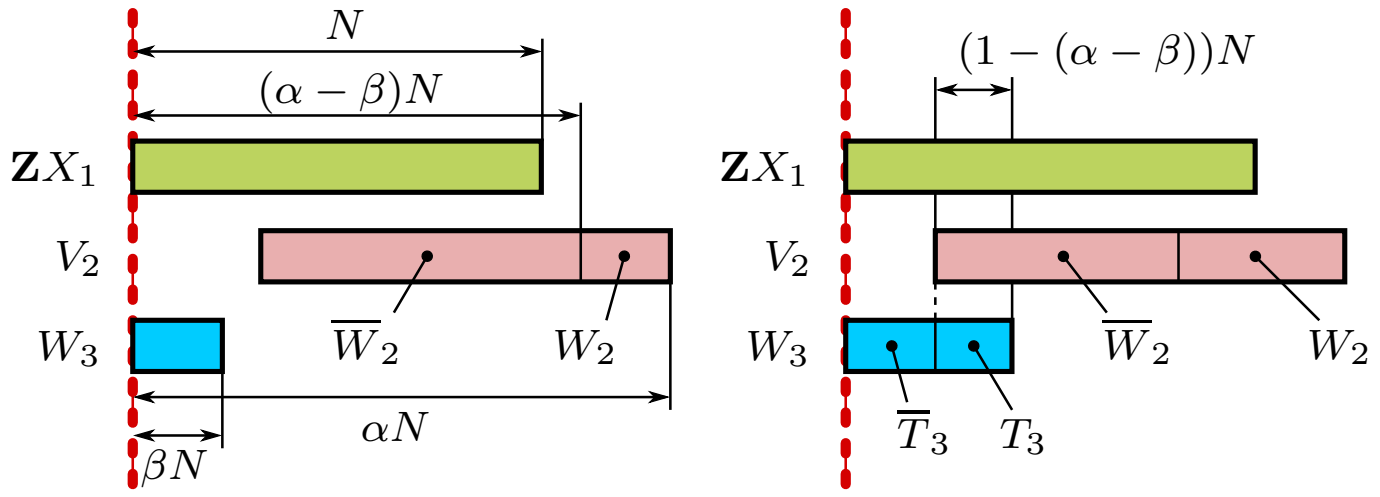

(a) Interferers do not overlap. (b) Interferers overlap.

**Fig. 3.** Components of received signal $Y_1$ for the converse proof. The components are shown sideways, with the bottom pipe on the left. The dotted vertical line symbolizes the "noise level", i.e., the lower end of the vector where further down-shifts cause loss of information. The received signal $Y_1$ is the modulo-2 sum of the three components.

Hence we focus on proving the bound $\mathcal{V}(\alpha - \beta)$ by generalizing the methods introduced in [10] to the case at hand. First note that Fano's inequality implies (with some abuse of notation for brevity) for every $k$

$$nR_k \leq I(X_k^n; Y_k^n).$$

### A. Without overlap between interferers

First consider $\alpha - \beta \geq 1$, which corresponds to the first line in the definition of $\mathcal{V}(\alpha - \beta)$. In this case, the two interfering signals do not overlap within the received signal, as shown in Fig. 3 (a). For example, at receiver 1, the sparsity patterns of $V_2$ and $W_3$ are disjoint. We can write

$$\begin{aligned} I(X_1^n; Y_1^n) &\overset{(a)}{=} I(X_1^n; Y_1^n W_2^n) \\ &= I(X_1^n; W_2^n) + I(X_1^n; Y_1^n \mid W_2^n) \\ &\overset{(b)}{=} H(Y_1^n \mid W_2^n) - H(Y_1^n \mid X_1^n, W_2^n), \end{aligned}$$

where (a) is with equality since $W_2$ is not interfered with in $Y_1$, and (b) uses the independence between $X_1$ and $W_2$. Now consider the last term.

$$\begin{aligned}H(Y_1^n \mid X_1^n, W_2^n) &= H(\mathbf{Z}X_1^n + V_2^n + W_3^n \mid X_1^n, W_2^n)\\ &= H(V_2^n + W_3^n \mid W_2^n)\\ &\overset{(a)}{=} H(W_3^n) + H(V_2^n \mid W_2^n)\\ &= H(W_3^n) + H(\overline{W}_2^n \mid W_2^n),\end{aligned}$$

where (a) follows from the fact that $V_2$ and $W_3$ do not overlap and different transmitters' signals are independent, and $\overline{W}_2$ is the part of $X_2$ that is not contained in $W_2$ (see Fig. 3 (a)). We conclude that

$$I(X_1^n; Y_1^n) = H(Y_1^n \mid W_2^n) - H(W_3^n) - H(\overline{W}_2^n \mid W_2^n).$$

Writing an analogous equation for $I(X_2^n; Y_2^n)$ and $I(X_3^n; Y_3^n)$, and adding all three of them, we arrive at

$$\begin{aligned}n \textstyle\sum_k R_k &\le H(Y_1^n \mid W_2^n) + H(Y_2^n \mid W_3^n) + H(Y_3^n \mid W_1^n)\\ &\quad - H(W_1^n) - H(\overline{W}_1^n \mid W_1^n) - H(W_2^n)\\ &\quad - H(\overline{W}_2^n \mid W_2^n) - H(W_3^n) - H(\overline{W}_3^n \mid W_3^n)\\ &= H(Y_1^n \mid W_2^n) + H(Y_2^n \mid W_3^n) + H(Y_3^n \mid W_1^n)\\ &\quad - H(X_1^n) - H(X_2^n) - H(X_3^n)\end{aligned}$$

Considering that $nR_k \le H(X_k^n)$, we conclude that

$$\begin{aligned}2n \textstyle\sum_k R_k &\le H(Y_1^n \mid W_2^n) + H(Y_2^n \mid W_3^n) + H(Y_3^n \mid W_1^n)\\ &\le nH(Y_1 \mid W_2) + nH(Y_2 \mid W_3) + nH(Y_3 \mid W_1),\end{aligned}$$

where single-letterization is performed by using the chain rule and omitting part of the conditioning. The right hand side of the last equation is maximized by letting each input bit pipe be independent $\mathrm{Bern}(1/2)$. Thus

$$\begin{aligned}2 \textstyle\sum_k R_k &\le 3N(\alpha - \beta), \qquad \text{and finally,}\\ d_{\text{sym}} &= \tfrac{R_{\text{sym}}}{N} \le \tfrac{\alpha-\beta}{2}.\end{aligned}$$

### *B. With overlap between interferers*

Now consider the case where $\alpha - \beta < 1$, i.e., the two interfering signals at each receiver overlap in signal space, see Fig. 3 (b). Define the top $(1 - (\alpha - \beta))N$ part of $X_k$ as $T_k$. We will augment the genie information $W_2^n$ of the previous subsection by $T_3^n$. This is exactly the part of the $X_3$-based interference that overlaps with the $X_2$-based interference.

Similar to the previous section, we conclude

$$\begin{aligned}I(X_1^n; Y_1^n) &= I(X_1^n; Y_1^n, W_2^n, T_3^n)\\ &= I(X_1^n; W_2^n, T_3^n) + I(X_1^n; Y_1^n \mid W_2^n, T_3^n)\\ &= H(Y_1^n \mid W_2^n, T_3^n) - H(Y_1^n \mid X_1^n, W_2^n, T_3^n),\end{aligned}$$

The last term becomes

$$\begin{aligned}H(Y_1^n \mid X_1^n, W_2^n, T_3^n) &= H(\mathbf{Z}X_1 + V_2^n + W_3^n \mid X_1^n, W_2^n, T_3^n)\\ &= H(V_2^n + W_3^n \mid W_2^n, T_3^n)\\ &\overset{(a)}{=} H(\overline{T}_3^n \mid T_3^n) + H(\overline{W}_2^n \mid W_2^n),\end{aligned}$$

where $\overline{T}_3$ denotes the part of $W_3$ that is not included in $T_3$. Its size is $N(\alpha - 1)$. We are allowed to separate the terms in (a) because the overlapping part is resolved by $T_3$.

Again, repeating the same for all three rates, we arrive at

$$\begin{aligned}n\textstyle\sum_k R_k &\le H(Y_1^n \mid W_2^n, T_3^n) - H(\overline{T}_1^n \mid T_1^n) - H(\overline{W}_1^n \mid W_1^n)\\ &\quad + H(Y_2^n \mid W_3^n, T_1^n) - H(\overline{T}_2^n \mid T_2^n) - H(\overline{W}_2^n \mid W_2^n)\\ &\quad + H(Y_3^n \mid W_1^n, T_2^n) - H(\overline{T}_3^n \mid T_3^n) - H(\overline{W}_3^n \mid W_3^n).\end{aligned}$$

Since $T_k$ and $\overline{T}_k$ together form $W_k$, which together with $\overline{W}_k$ forms $X_k$, we can write

$$nR_1 \le H(X_1^n) = H(T_1^n) + H(\overline{T}_1^n | T_1^n) + H(\overline{W}_1 | \underbrace{T_1^n, \overline{T}_1^n}_{=W_1^n})$$

Using this expression and its equivalent for $R_2$ and $R_3$ with the previous inequality, we obtain

$$\begin{aligned}2n\textstyle\sum_k R_k &\le H(Y_1^n \mid W_2^n, T_3^n) + H(T_1^n) + H(Y_2^n \mid W_3^n, T_1^n)\\ &\quad + H(T_2^n) + H(Y_3^n \mid W_1^n, T_2^n) + H(T_3^n)\\ &\le n\big(H(Y_1 \mid W_2, T_3) + H(T_1) + H(Y_2 \mid W_3, T_1)\\ &\quad + H(T_2) + H(Y_3 \mid W_1, T_2) + H(T_3)\big).\end{aligned}$$

Again, the right hand side is maximized by choosing all $X_k$ components independently according to $\mathrm{Bern}(1/2)$, yielding

$$\begin{aligned}2 \textstyle\sum_k R_k &\le 3N + 3N(1 - (\alpha - \beta)),\\ d_{\text{sym}} &\le 1 - \tfrac{\alpha-\beta}{2},\end{aligned}$$

which matches the definition of $\mathcal{V}(\alpha - \beta)$ for $\alpha - \beta < 1$.

## V. Achievability proof

The set of interest $\{(\alpha, \beta)\}$ is divided into regions "Aa" to "Df" as shown in Fig. 2. In each region, we use the following coding scheme. For every sender $k$, we set

$$X_k = \mathbf{G}D_k,$$

where $\mathbf{G} \in \mathbb{F}_2^{N \times d_{\text{sym}}N}$ is the *assignment matrix*, and $D_k \in \mathbb{F}_2^{d_{\text{sym}}N}$ is a vector of i.i.d. $\mathrm{Bern}(1/2)$ message bits.

We constrain the coding scheme in several ways, namely, (a) there is no coding across multiple channel uses, (b) all transmitters use the same $\mathbf{G}$, and (c) the proposed $\mathbf{G}$ matrices will have at most one non-zero element per row, i.e., each pipe in $X_k$ is assigned either an information bit or a zero. While these assumptions may seem overly restrictive, they are sufficient for our purposes. Indeed, it is surprising that such a constrained set of codes is able to meet the upper bound of Section IV.

*Remark 3:* If the number of input pipes $N$ is small, it can severely limit our options in terms of assignment matrices. The following argument can circumvent this problem by expanding a given channel to one with more input pipes. To this end, consider $L \ge 2$ subsequent channel uses, with channel inputs $X_{k,1}, \ldots, X_{k,L}$. By interleaving these vectors into a supersymbol $\widetilde{X}_k = \sum_{l=1}^{L} (\mathbf{I}_N \otimes \mathbf{e}_l) X_{k,l}$, and likewise for the outputs $\widetilde{Y}_k$, it can be shown that the resulting channel $\{\widetilde{X}_1, \widetilde{X}_2, \widetilde{X}_3\} \to \{\widetilde{Y}_1, \widetilde{Y}_2, \widetilde{Y}_3\}$ is in fact $(LN, \alpha, \beta)$ as defined in Section II. (Here, $\otimes$ denotes the Kronecker product, and $\mathbf{e}_l$ is the $l$th column of $\mathbf{I}_L$.) Through this method, a channel with a given $N$ can be expanded to one with $LN$ input pipes. Note that $d_{\text{sym}}$ is unaffected by this transformation since it is normalized by the number of pipes. In light of this transformation, we assume from now on that $N$ is (or has been made) large enough such that any fraction of $N$ that we incur corresponds to an integer number of pipes.

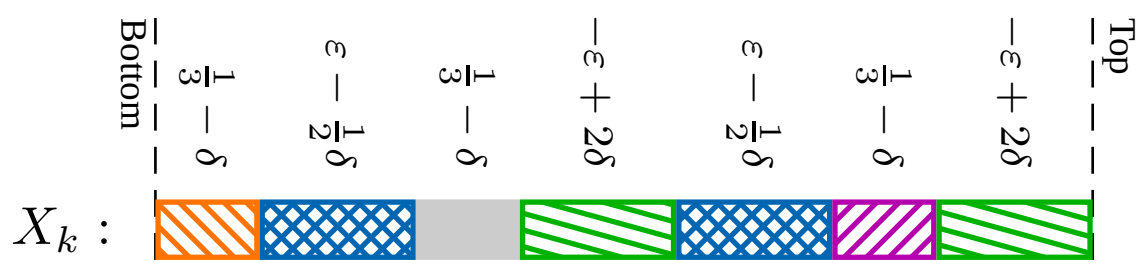

**Fig. 4.** Proposed $\mathbf{G}$ assignment for region "Df", expressed in terms of the transmit vector. The block lengths are given as fractions of $N$, such that the sum of all block lengths is 1.

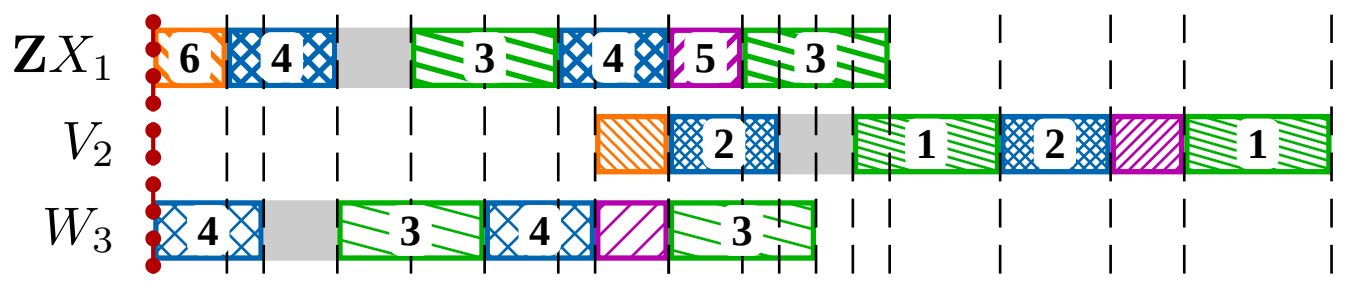

**Fig. 5.** Received signal $Y_1$ in "Df", at $\alpha = 1.6$, $\beta = 0.9$, with $d_{\text{sym}} = 0.55$. Blocks in different rows carry different data.

Optimal assignment matrices $\mathbf{G}$ for all regions in Fig. 2 are listed in Table 1. An interactive online animation is also available at [11]. Each row in the table contains the definition of a region in terms of affine constraints in $(\alpha, \beta)$ and a representation of $\mathbf{G}$ by means of the resulting transmit vector $X_k$. In the following we discuss the details for one particular example, which is representative for all other cases.

*Example (Region "Df"):* This region is parameterized by $(\alpha, \beta) = (4/3 + \varepsilon,\ 2/3 + \delta)$ with $\varepsilon \leq 2\delta,\ \varepsilon \geq \frac{1}{2}\delta,\ \delta \leq \frac{1}{3}$. Fig. 4, copied from Table 1, represents an optimal assignment $\mathbf{G}$ by means of the resulting transmit vector. The vector $X_k$ is subdivided into *data blocks* (hatched) that correspond to non-zero rows of $\mathbf{G}$, and *zero blocks* (gray) that correspond to all-zero rows of $\mathbf{G}$. Some data blocks occur twice. We denote such block pairs as *twins*. Twins carry the same data bits, albeit in reverse order as discussed later. The length of each block as a fraction of $N$ is annotated in the figure.

To prove achievability of Theorem 1, we require the transmit vector to be both *valid* and *decodable*. By valid we mean (a) all block lengths are non-negative for the range of $(\varepsilon, \delta)$ that constitute the region, (b) the sum of the block lengths is 1, and (c) adding the sizes of all data blocks, counting twins only once, results in the desired $d_{\text{sym}}$ as claimed in Theorem 1 ($2/3 - \delta/2$ in our example). By decodable, we mean that using this transmit vector assignment, the receiver can recover all desired data blocks from the received signal.

To verify decodability, consider Fig. 5, which uses the same conventions as Fig. 3. The receiver sees the sum of data blocks from different transmitters, each characterized by its length and shift location. Different blocks may or may not overlap. Decoding is performed sequentially, block by block. In each step, one of three rules is applied in order to decode additional data blocks, which are then removed from the received signal. The three decoding rules are as follows.

*1. Direct readout:*
Consider the situation in Fig. 6 (b). If a data block (i) does not overlap with any other data block and (ii) is located above the noise level, then its data content can be read out directly from the received signal.[1] A block that has been read out is then removed from the received signal. If the block has a twin, it is removed as well.

*2. Overlapping twins scenario (A):*
Consider Fig. 6 (c). If two twin pairs exist such that (i) they have the same block length, $b_1 = b_2$, (ii) they have the same separation, $s_1 = s_2$, (iii) the relative shift between the pairs is less than the separation, $c < s_1$, and (iv) the dashed sections of (A) in Fig. 6 (c) do not overlap with any other data block and are above the noise floor, then both twin pairs can be decoded and canceled from the received signal.[2] To see this, consider the following successive decoding argument [6]. Let the two copies within a twin be in reverse order of each other. First, the leftmost part of the left blue twin is read out. Its data reappears on the right side of the right blue twin, thus revealing a chunk of data on the right side of the right yellow twin. This data in turn is replicated on the left side of the left yellow twin, which exposes a new part of the left blue twin. The process repeats until both twins are completely decoded.

*3. Overlapping twins scenario (B):*
This rule is a variant of the previous one, where pattern (B) replaces pattern (A) in Fig. 6 (c).

In our example, the sequence of steps that completely decodes $X_1$ is annotated in Fig. 5: First, block 1 is decoded via direct readout (rule 1). The now-known data block and its twin are removed from the received signal $Y_1$. The same rule allows block 2 to be decoded, which is then removed from $Y_1$. Each removal step makes more room for subsequent rule applications. Next, rule 2 is applied to the two pairs of twins 3. Continuing in the same fashion, the removal of blocks 1, 2 and 3 enables the two twin pairs 4 to be decoded using rule 3. Finally, data blocks 5 and 6 can be recovered by direct readout (rule 1), which completes the decoding process. By symmetry, the signals at the other two receivers can be similarly decoded.

[1] It is crucial that both (i) and (ii) hold for all $(\alpha, \beta)$ in the region, since the length and location of the blocks in Fig. 5 change when $\alpha$ and $\beta$ vary.

[2] Again, conditions (i)–(iv) must hold for all $(\alpha, \beta)$ in the region.

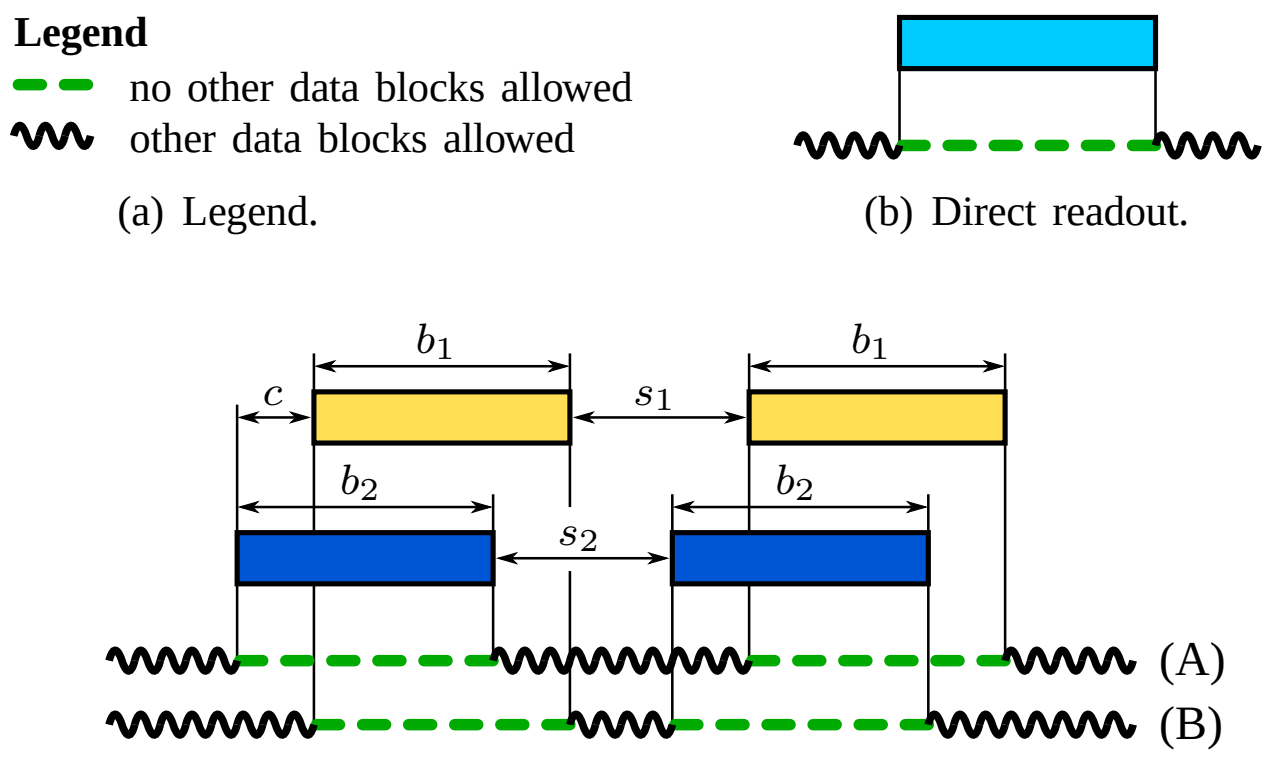

(a) Legend. (b) Direct readout. (c) Overlapping twins.

**Fig. 6.** Rules for verifying decodability. Legend (a) applies to "direct readout", shown in (b), and two variants of "overlapping twins", shown in (c).

The assignments for all other regions as listed in Table 1 can be shown to be valid and decodable using the same procedure.

| | $(\alpha, \beta)$ | Region constraints | $d_{\text{sym}}$ | Assignment (shown graphically and as corresponding list of block lengths) |
|---|---|---|---|---|
| Aa | $(2+\varepsilon, \delta)$ | $\delta \geq 0,\ \delta \leq 1+\varepsilon,\ \varepsilon \leq -\delta$ | $1+\frac{1}{2}\varepsilon-\frac{1}{2}\delta$ | $(\delta \mid -\frac{1}{2}\varepsilon-\frac{1}{2}\delta \mid 1+\varepsilon \mid -\frac{1}{2}\varepsilon-\frac{1}{2}\delta)$ |
| Ab | $(2+\varepsilon, \delta)$ | $\varepsilon \leq 0,\ \delta \leq \frac{1}{2},\ \varepsilon \geq -\delta$ | $1-\delta$ | $(\delta \mid 1-\delta)$ |
| Ba | $(6/5+\varepsilon, 2/5+\delta)$ | $\varepsilon \geq 3\delta,\ \varepsilon \leq \frac{1}{5}+\delta,\ \varepsilon \geq \frac{1}{10}-\frac{1}{2}\delta$ | $\frac{3}{5}-\frac{1}{2}\varepsilon+\frac{1}{2}\delta$ | $(\frac{1}{5}-\varepsilon+\delta \mid \frac{1}{5}-\frac{1}{2}\varepsilon-\frac{1}{2}\delta \mid \frac{1}{5}-\frac{1}{2}\varepsilon-\frac{1}{2}\delta \mid -\frac{1}{5}+2\varepsilon+\delta \mid \frac{1}{5}-\frac{1}{2}\varepsilon-\frac{1}{2}\delta \mid \frac{1}{5}-\frac{1}{2}\varepsilon-\frac{1}{2}\delta \mid \frac{1}{5}+\varepsilon)$ |
| Bb | $(6/5+\varepsilon, 2/5+\delta)$ | $\varepsilon \geq -\frac{1}{3}\delta,\ \varepsilon \leq \frac{1}{5}+\delta,\ \varepsilon \leq \frac{1}{10}-\frac{1}{2}\delta,\ \varepsilon \geq 3\delta$ | $\frac{3}{5}-\frac{1}{2}\varepsilon+\frac{1}{2}\delta$ | $(\frac{1}{5}-\varepsilon+\delta \mid \frac{1}{5}-\frac{1}{2}\varepsilon-\frac{1}{2}\delta \mid \frac{3}{2}\varepsilon+\frac{1}{2}\delta \mid \frac{1}{5}-2\varepsilon-\delta \mid \frac{3}{2}\varepsilon+\frac{1}{2}\delta \mid \frac{1}{5}-\frac{1}{2}\varepsilon-\frac{1}{2}\delta \mid \frac{1}{5}+\varepsilon)$ |
| Bc | $(6/5+\varepsilon, 2/5+\delta)$ | $\varepsilon \geq \frac{1}{2}\delta,\ \varepsilon \leq \frac{1}{5}+\delta,\ \varepsilon \leq -\frac{1}{3}\delta$ | $\frac{3}{5}-\frac{1}{2}\varepsilon+\frac{1}{2}\delta$ | $(\frac{1}{5}+\frac{1}{2}\delta \mid \frac{1}{5}+\varepsilon \mid -\frac{3}{2}\varepsilon-\frac{1}{2}\delta \mid \frac{1}{5}+\varepsilon \mid -\frac{3}{2}\varepsilon-\frac{1}{2}\delta \mid \frac{1}{5}+\varepsilon \mid \frac{1}{5}+\frac{1}{2}\delta)$ |
| Bd | $(6/5+\varepsilon, 2/5+\delta)$ | $\varepsilon \leq \frac{1}{2}\delta,\ \varepsilon \leq -2\delta,\ \varepsilon \geq -\frac{1}{5}$ | $\frac{3}{5}+\frac{1}{2}\varepsilon$ | $(-\varepsilon+\frac{1}{2}\delta \mid \frac{1}{5}+\varepsilon \mid -\varepsilon+\frac{1}{2}\delta \mid \frac{1}{5}+\varepsilon \mid -\frac{1}{2}\varepsilon-\delta \mid \frac{1}{5}+\varepsilon \mid -\frac{1}{2}\varepsilon-\delta \mid \frac{1}{5}+\varepsilon \mid -\varepsilon+\frac{1}{2}\delta \mid \frac{1}{5}+\varepsilon \mid -\varepsilon+\frac{1}{2}\delta)$ |
| Be | $(6/5+\varepsilon, 2/5+\delta)$ | $\varepsilon \leq \frac{1}{2}\delta,\ \varepsilon \geq -2\delta,\ \delta \leq \frac{1}{10}$ | $\frac{3}{5}-\delta$ | $(-\varepsilon+\frac{1}{2}\delta \mid \frac{1}{5}+\varepsilon \mid -\varepsilon+\frac{1}{2}\delta \mid \frac{1}{5}+\varepsilon \mid \frac{1}{5}-2\delta \mid \frac{1}{5}+\varepsilon \mid -\varepsilon+\frac{1}{2}\delta \mid \frac{1}{5}+\varepsilon \mid -\varepsilon+\frac{1}{2}\delta)$ |
| Bf | $(6/5+\varepsilon, 2/5+\delta)$ | $\varepsilon \geq \frac{1}{2}\delta,\ \varepsilon \leq 3\delta,\ \varepsilon \leq \frac{1}{10}-\frac{1}{2}\delta$ | $\frac{3}{5}-\delta$ | $(\frac{1}{5}-\varepsilon+\delta \mid \frac{1}{5}-\varepsilon+\delta \mid 2\varepsilon-\delta \mid \frac{1}{5}-2\varepsilon-\delta \mid 2\varepsilon-\delta \mid \frac{1}{5}-\varepsilon+\delta \mid \frac{1}{5}+\varepsilon)$ |
| Bg | $(6/5+\varepsilon, 2/5+\delta)$ | $\varepsilon \leq 3\delta,\ \delta \leq \frac{1}{10},\ \varepsilon \geq \frac{1}{10}-\frac{1}{2}\delta$ | $\frac{3}{5}-\delta$ | $(\frac{1}{5}-\varepsilon+\delta \mid \frac{1}{5}-\varepsilon+\delta \mid \frac{1}{5}-2\delta \mid -\frac{1}{5}+2\varepsilon+\delta \mid \frac{1}{5}-2\delta \mid \frac{1}{5}-\varepsilon+\delta \mid \frac{1}{5}+\varepsilon)$ |
| Da | $(4/3+\varepsilon, 2/3+\delta)$ | $\varepsilon \geq 2\delta,\ \varepsilon \geq -\delta,\ \varepsilon \leq \frac{1}{3}+\delta$ | $\frac{2}{3}-\frac{1}{2}\varepsilon+\frac{1}{2}\delta$ | $(\frac{1}{3}-\delta \mid \frac{1}{2}\varepsilon+\frac{1}{2}\delta \mid \frac{1}{3}-\delta \mid \frac{1}{2}\varepsilon+\frac{1}{2}\delta \mid \frac{1}{3}-\varepsilon+\delta)$ |
| Db | $(4/3+\varepsilon, 2/3+\delta)$ | $\varepsilon \leq -\delta,\ \varepsilon \geq \frac{1}{2}\delta,\ \delta \geq -\frac{1}{6}$ | $\frac{2}{3}+\delta$ | $(\frac{1}{3}+\frac{1}{2}\delta \mid \frac{1}{3}-\delta \mid \frac{1}{3}+\frac{1}{2}\delta)$ |
| Dc | $(4/3+\varepsilon, 2/3+\delta)$ | $\varepsilon \leq \frac{1}{2}\delta,\ \varepsilon \geq 2\delta,\ \delta \geq -\frac{1}{6}$ | $\frac{2}{3}+\delta$ | $(-\varepsilon+\frac{1}{2}\delta \mid \frac{1}{3}+\varepsilon \mid -\varepsilon+\frac{1}{2}\delta \mid \frac{1}{3}+2\varepsilon-2\delta \mid -\varepsilon+\frac{1}{2}\delta \mid \frac{1}{3}+\varepsilon \mid -\varepsilon+\frac{1}{2}\delta)$ |
| Df | $(4/3+\varepsilon, 2/3+\delta)$ | $\varepsilon \leq 2\delta,\ \varepsilon \geq \frac{1}{2}\delta,\ \delta \leq \frac{1}{3}$ | $\frac{2}{3}-\frac{1}{2}\delta$ | $(\frac{1}{3}-\delta \mid \varepsilon-\frac{1}{2}\delta \mid \frac{1}{3}-\delta \mid -\varepsilon+2\delta \mid \varepsilon-\frac{1}{2}\delta \mid \frac{1}{3}-\delta \mid -\varepsilon+2\delta)$ |
| Ea | $(2+\varepsilon, 2/3+\delta)$ | $\varepsilon \leq 0,\ \delta \geq -\frac{1}{15},\ \varepsilon \geq 3\delta$ | $\frac{2}{3}+\delta$ | $(-3\delta \mid \frac{1}{3}+2\delta \mid -3\delta \mid \frac{1}{3}+5\delta \mid -3\delta \mid \frac{1}{3}+2\delta)$ |
| Eb | $(2+\varepsilon, 2/3+\delta)$ | $\varepsilon \leq 0,\ \delta \leq -\frac{1}{15},\ \delta \geq -\frac{1}{6},\ \varepsilon \geq 3\delta$ | $\frac{2}{3}+\delta$ | $(-3\delta \mid \frac{1}{3}+2\delta \mid \frac{1}{3}+2\delta \mid -\frac{1}{3}-5\delta \mid \frac{1}{3}+2\delta \mid \frac{1}{3}+2\delta)$ |
| Ec | $(2+\varepsilon, 2/3+\delta)$ | $\varepsilon \leq 3\delta,\ \varepsilon \geq -\frac{1}{3}+\delta,\ \varepsilon \leq -\frac{1}{3}-2\delta$ | $\frac{2}{3}+\frac{1}{2}\varepsilon-\frac{1}{2}\delta$ | $(-\frac{1}{2}\varepsilon-\frac{3}{2}\delta \mid \frac{1}{3}+\frac{1}{2}\varepsilon+\frac{1}{2}\delta \mid \frac{1}{3}+\frac{1}{2}\varepsilon+\frac{1}{2}\delta \mid -\frac{1}{3}-\varepsilon-2\delta \mid \frac{1}{3}+\frac{1}{2}\varepsilon+\frac{1}{2}\delta \mid \frac{1}{3}+\frac{1}{2}\varepsilon+\frac{1}{2}\delta \mid -\frac{1}{2}\varepsilon+\frac{3}{2}\delta)$ |
| Ed | $(2+\varepsilon, 2/3+\delta)$ | $\varepsilon \leq 3\delta,\ \varepsilon \geq -\frac{1}{3}-2\delta,\ \varepsilon \geq -\frac{1}{3}+\delta,\ \varepsilon \leq -3\delta$ | $\frac{2}{3}+\frac{1}{2}\varepsilon-\frac{1}{2}\delta$ | $(-\frac{1}{2}\varepsilon-\frac{3}{2}\delta \mid \frac{1}{3}+\frac{1}{2}\varepsilon+\frac{1}{2}\delta \mid -\frac{1}{2}\varepsilon-\frac{3}{2}\delta \mid \frac{1}{3}+\varepsilon+2\delta \mid -\frac{1}{2}\varepsilon-\frac{3}{2}\delta \mid \frac{1}{3}+\frac{1}{2}\varepsilon+\frac{1}{2}\delta \mid -\frac{1}{2}\varepsilon+\frac{3}{2}\delta)$ |
| Ee | $(2+\varepsilon, 2/3+\delta)$ | $\varepsilon \leq 0,\ \delta \leq \frac{1}{3}+\varepsilon,\ \delta \geq -\frac{1}{3}\varepsilon$ | $\frac{2}{3}+\frac{1}{2}\varepsilon-\frac{1}{2}\delta$ | $(\frac{1}{3}-\delta \mid \frac{1}{3}-\delta \mid \frac{1}{2}\varepsilon+\frac{3}{2}\delta \mid \frac{1}{3}-\delta \mid \frac{1}{2}\varepsilon+\frac{3}{2}\delta \mid -\varepsilon)$ |

**Table 1.** Assignments that achieve $d_{\text{sym}}$ as stated in Theorem 1.